\documentstyle[11pt,appb,epsf]{article}

\begin{document}
\makeatletter

\def\l@plain{lplain}
\newif\ifLaTeX@ld
\ifx\fmtname\l@plain \LaTeX@ldtrue \else \LaTeX@ldfalse \fi

\font\appheadcs=cmcsc10 at 10 truept
\font\appheadit=cmti10 at 10 truept
\font\appauthcs=cmcsc10 scaled 1095 
\font\titlefont=cmr7 scaled 1728
\renewcommand{\baselinestretch}{0.94}

\def\section{\@startsection {section}{1}{\z@}{3.5ex plus-1ex minus
    -.2ex}{0.1ex plus.1ex}{\reset@font\normalsize\bf}}
\def\subsection{\@startsection{subsection}{2}{\z@}{3.25ex plus-1ex
     minus-.2ex}{1ex plus.2ex}{\reset@font\normalsize\it}}
\def\subsubsection{\@startsection{subsubsection}{3}{\z@}{-3.25ex plus
     -1ex minus-.2ex}{1.5ex plus.2ex}{\reset@font\normalsize\it}}
\def\paragraph{\@startsection
     {paragraph}{4}{\z@}{3.25ex plus1ex minus.2ex}{-1em}{\reset@font
     \normalsize\bf}}
\def\subparagraph{\@startsection
     {subparagraph}{4}{\parindent}{3.25ex plus1ex minus
     .2ex}{-1em}{\reset@font\normalsize\bf}}
\makeatother

\null
\hfill WUE-ITP-98-008\\
\null
\hfill hep-ph/9803304\\
\vskip .8cm
\begin{center}
{\Large \bf Spin Effects in Neutralino Production\\[.5em]
in $e^+e^-$ Annihilation with\\[.5em]
polarized beams%
\footnote{Work supported by the German Federal Ministry for
Research and Technology (BMBF) under contract number
05 7WZ91P (0).}
}
\vskip 2.5em

{\large \sf G. Moortgat-Pick$^1$,
H. Fraas$^1$, A. Bartl$^2$, W. Majerotto$^3$},
\address{ 
$^1$ Institute of Theoretical Physics,
University of W\"urzburg, Germany,\\
$^2$ Institute of Theoretical Physics,
University of Vienna, Austria,\\
$^3$ Institute of High Energy Physics, Austrian
 Academy of Sciences, Vienna, Austria}
\end{center}

\begin{abstract}
We study
the process $e^+ e^- \to\tilde{\chi}^0_1\tilde{\chi}^0_2$
 and the subsequent
decay $\tilde{\chi}^0_2 \to \tilde{\chi}^0_1 \ell^{+}\ell^{-}$
 with polarized beams, including the complete spin correlations of $\tilde{\chi}^0_2$.
We present numerical results for the
 lepton energy and angular distributions, and for the
distribution of the opening angle between the leptons
 for $\sqrt{s}=500$ GeV. 
The effects of spin correlations are important in the lepton angular distribution,
especially for gaugino--like neutralinos.
 The opening angle distribution is
particularly suitable for distinguishing between higgsino- and gaugino-like neutralinos.
The polarized cross sections 
and lepton angular distributions are very sensitive to the mixing of the
neutralinos.
\end{abstract}

\null
\setcounter{page}{0}
\clearpage
%

\section{Introduction}
\vspace{-.3cm}
Most studies of neutralino production $e^+e^- \to \tilde{\chi}^0_i 
\tilde{\chi}^0_j$ and decay of neutralinos have been performed in the 
Minimal Supersymmetric Standard Model (MSSM). (See, for example 
\cite{bartl}, \cite{ambrosiano}, \cite{desy}, and references therein.)
Usually, in these studies the spin correlations between the
production and decay process have been neglected. It is expected that
these spin correlations are important in the angular distributions.

Angular
 distributions and angular correlations of the decay products of
neutralinos can give
valuable information on their mixing character. Their measurement is
 very suitable for constraining the parameter space of the MSSM. In
 this contribution we study $e^+e^- \to \tilde{\chi}^0_1
 \tilde{\chi}^0_2$ with the subsequent decay $\tilde{\chi}^0_2 \to
\tilde{\chi}^0_1 \ell^+ \ell^-$, including the complete spin correlations of
$\tilde{\chi}^0_2$. We also take into account beam polarization. We
 give numerical results for polarized cross sections,
 forward--backward asymmetries, lepton angular
distributions, lepton opening angular distributions and lepton energy 
distributions at $\sqrt{s} = 500$~GeV.
\section{General Formalism}
\vspace{-.3cm}
 Both the production process,
$e^{+} e^{-} \to \tilde{\chi}^0_1 \tilde{\chi}^0_2$, and
the decay process, $\tilde{\chi}^0_2 \to
\tilde{\chi}^0_1 \ell^{+} \ell^{-}$
 contain contributions from $Z^0$ exchange and from $\tilde{\ell}_L$ and $\tilde{\ell}_R$
 exchange. A treatment of these processes properly taking into account
 the polarization
of $\tilde{\chi}^0_2$ has been given in \cite{moor} following the method of
\cite{haber}.

The amplitude $T=\Delta_2 P^{\lambda_2} D_{\lambda_2}$
 of the combined process is
a product of the helicity amplitude
$P^{\lambda_2}$ for the production process times the helicity amplitude
 $D_{\lambda_2}$ for the decay process and the propagator
$\Delta_2=1/[s_2-m_2^2+i m_2 \Gamma_2]^{-1}$ of $\tilde{\chi}^0_2$ with helicity
$\lambda_2$, summed over $\lambda_2$.
$s_2$, $m_2$, $\Gamma_2$ denote the four--momentum squared, mass and width of
$\tilde{\chi}^0_2$. The amplitude squared
$|T|^2=|\Delta_2|^2 \rho^{\lambda_2 \lambda_2'}_P
\rho^D_{\lambda_2' \lambda_2} \label{N}$
 is thus composed of the
unnormalized spin density production matrix
 $\rho^{\lambda_2 \lambda_2'}_P=P^{\lambda_2} P^{\lambda_2' *}$
of $\tilde{\chi}^0_2$ and
 the decay matrix
 $\rho^{D}_{\lambda_2' \lambda_2}=D_{\lambda_2} D_{\lambda_2'}^{*}$. 

 Interference terms between
various helicity amplitudes preclude factorization in a production
factor  $\sum_{\lambda_2} |P^{\lambda_2}|^2$ times a decay factor
$\overline{\sum}_{\lambda_2} |D_{\lambda_2}|^2$. The analytical
formulae are given in \cite{moor}.

\section{Numerical Results and Discussion}
\vspace{-.3cm}
Neutralinos are linear superpositions 
 of the photino $\tilde{\gamma}$ and
the zino $\tilde{Z}$ and the two
higgsinos $\tilde{H}^0_a$ and $\tilde{H}^0_b$.
The composition of the neutralino states depend on the three
SUSY mass parameters $M, M'$ (with the GUT relation $M'=\frac{5}{3} M
\tan^2\theta_W$)
 and $\mu$, and on the ratio $\tan\beta=v_2/v_1$ of the vacuum expectation
values of the Higgs fields \cite{2haber}.
 The masses of the sleptons are determined by the common
scalar mass parameter $m_0$, and $M$ and
 $\tan\beta$ \cite{hall}.
 We shall consider two representative scenarios which differ significantly
in the nature of the two lowest mass eigenstates $\tilde{\chi}^0_1$
and $\tilde{\chi}^0_2$: (A) M=380 GeV, $\mu=-180$ GeV, $\tan\beta=2$,
$m_0=80$ GeV;
 (B) M=185 GeV, $\mu=550$ GeV, $\tan\beta=1.6$, $m_0=280$ GeV. Using the short--hand
notation $\tilde{\chi}^0_i = (\tilde{\gamma}|\tilde{Z}|\tilde{H}^0_a|
\tilde{H}^0_b)$, in (A) both neutralinos
$\tilde{\chi}^0_1=(-.37|+.29|-.21|-.86), m_{\tilde{\chi}^0_1}=170$ GeV and
$\tilde{\chi}^0_2=(-.03|+.17|-.94|-.30), m_{\tilde{\chi}^0_2}=195$ GeV
 have strong higgsino components. In
 (B) $\tilde{\chi}^0_1=(+.83|-.55|+.09|+.04), m_{\tilde{\chi}^0_1}=88$
 GeV and $\tilde{\chi}^0_2=(-.56|-.81|+.18|+.06),
 m_{\tilde{\chi}^0_2}=170$ GeV
 have dominating gaugino components,
$\tilde{\chi}^0_1$ being almost a pure bino and $\tilde{\chi}^0_2$
nearly a pure wino. 
\subsection{ Unpolarized beams}
\vspace{-.5cm}
The total cross section for the combined process is independent of
spin correlations \cite{dicus}. For unpolarized beams it is 4.7~fb in scenario
(A) and 4.1~fb in scenario (B).
\vspace{-.8cm}
\subsubsection{Lepton angular distributions}
\vspace{-.5cm}
In Figs.~1 and 2 we present numerical results for the distribution
 $d\sigma/d\cos\Theta_{-}$ with $\Theta_-$ being the angle between
the outgoing leptons $\ell^{-}$ and the electron beam in the
laboratory system. We compare our results
with those assuming factorization of the
differential cross section into production and decay (Figs.~1--2).

The spin effect is largest in the forward and backward direction.
For
gaugino-like neutralinos it amounts to about $15\%$, for higgsino--like neutralinos it is
smaller.
 The resulting forward-backward asymmetry $A_{FB}$ depends sensitively on the
mixing character of the neutralinos. For higgsino-like neutralinos (scen.~(A)) the angular
distribution is nearly forward--backward symmetric with a minimum at $\Theta_- = 90^o$
 For gaugino-like neutralinos (scen.~(B)), however,
the forward hemisphere is favoured (Table~1).

\vspace{-.5cm}
\subsubsection{The lepton opening angle distributions}
\vspace{-.5cm}
\begin{sloppypar}
Owing to the Majorana character of the neutralinos
the distribution $d\sigma/d\cos\Theta_{+-}$
 of the opening angle between both outgoing leptons
factorizes.
\end{sloppypar}

The distributions are completely different for
scenario (A). (Fig.~6) and (B) (Fig.~7).
For gaugino-like neutralinos it is rather flat with a maximum near
$\Theta_{+-}=40^0$, whereas for higgsino--like neutralinos it is much steeper with a
peak at $\Theta_{+-}=0^0$ (notice the different scales in Figs.~6 and
7). For gaugino-like neutralinos the position of the maximum depends
strongly on
the kinetic energy of the decaying neutralino.
With increasing kinetic energy it moves from $\pi/2$ to $0^0$. For
higgsino-like neutralinos the maximum is always at $0^0$, but the
distribution becomes much steeper with increasing energy.
In both scenarios most of the leptons are emitted with an opening
angle between 0 and $\pi/2$ (approximately $76\%$ for scen.~(A) and
$69\%$ for scen.~(B)). The opening angle distribution depends only slightly on the
value of
$m_0$. Therefore, this distribution is more suitable for the
discrimination between gaugino- and higgsino-like neutralinos than the lepton angular
distribution which has a stronger $m_0$ dependence (compare with \cite{2moor}).

\vspace{-.5cm}
\subsection{Polarized beams}
\vspace{-.5cm}
We have computed the angular and energy distributions for longitudinal
polarization $P_{-}=\pm 0.9$ of the electron beam and
$P_{+}=\pm 0.4$ of the positron beam
( $P_{\pm}>0$ ($P_{\pm}<0$) for right-handed (left-handed) polarized
beams).

In the following the different polarization states ared denoted by the sign of $P_{-}$ and
$P_{+}$, (0 0) denotes the case of unpolarized
beams.

In Table~1 the total cross sections for the different
beam polarizations are given in
increasing order of magnitude. For higgsino-like
 neutralinos the cross section is highest for right-handed
 electrons and left-handed positrons. For the gaugino case it is highest for left-handed
electrons and right-handed positrons due to
 the wino-like character of $\tilde{\chi}^0_2$. For gaugino-like
 neutralinos the dependence on the beam polarization is also more pronounced
 than for the higgsino case.
\vspace{-.5cm}
\begin{center}
\begin{table}
\begin{tabular}{|l|c|c|c|c|c|c|c|}\hline
A & $(--)$ & (++) & $(-0)$ & (00) & (+0) & $(-+)$ & $(+-)$ \\ \hline
$\sigma$ /fb & 2.9 & 3.2 & 4.5 & 4.7 & 5.0 & 6.1 & 6.8 \\ \hline
$A_{FB}$ / \% & .48 & -.75 & .58 & -.16 & -.83 & .63 & -.87 \\ \hline\hline
B & (++) & (+0) & $(+-)$ & (00) & $(--)$ & $(-0)$ & $(-+)$ \\ \hline
$\sigma$ /fb & .87 & .94 & 1.0 & 4.1 & 4.4 & 7.2 & 10.1 \\ \hline
$A_{FB}$ / \% & 2.6 & -1.8 & -5.6 & 9.2 & 11 & 11 & 11\\ \hline
\end{tabular}
\vspace{.3cm}
\caption{Total cross sections and Asymmetries for different
  combinations (sign$P_{-}$,sign$P_{+}$) of electron polarization
  $P_{-}=\pm .9$ and positron polarization $P_{+}=\pm .4$}
\end{table}
\end{center}
\vspace{-1cm}
\subsubsection{Lepton angular distribution for polarized beams}
\vspace{-.5cm}
For higgsino-like neutralinos the distribution
 is nearly independent of the beam polarization and almost the same
 as for unpolarized  beams (Fig.~3). Similarly, for the gaugino-like
 scenario (B) and left-handed electrons only the magnitude of the
 distribution changes for different positron polarization (Fig.~4).
 The consequence of non-factorization of the angular distribution in
 production and decay is most pronounced for gaugino-like
 neutralinos and right-handed electrons (Fig.~5).

For unpolarized beams the differential cross section $d\sigma /\cos\Theta_-$ is larger in
the forward hemisphere. For
$e^{-}e^{+}$-polarization (+0) the distribution is nearly symmetric.
For beam polarization $(+-)$ the backward hemisphere is favoured.
\vspace{-.5cm}
\subsubsection{The lepton opening angle distribution for polarized
  beams}
\vspace{-.5cm}
Also in the case of polarized beams the opening angle
distribution factorizes into the contributions from production and
decay due to the Majorana nature of the neutralino. In both scenarios the only effect of
beam polarization is to increase or to reduce the cross sections without changing the
shape. The dependence on beam polarization 
is the same as that for the total cross section, Table~1. It is
different for gaugino- and higgsino-like neutralinos. Figs.~(6$-$7) shows
the opening angle distribution for different beam polarization for
higgsino-like neutralinos, scen.~(A) and for
gaugino-like neutralinos, scen.~(B).
\vspace{-.5cm}
\subsubsection{Energy distributions for polarized beams}
\vspace{-.5cm}
The energy distributions of
the outgoing leptons in the laboratory frame again factorize due to the
Majorana character of the decaying $\tilde{\chi}^0_2$. Since in both
scenarios the shape is very similar, we only show the energy spectra
of $\ell^{-}$ for scen.~(B) and different combinations of beam
polarization (Fig.~8). As a consequence of factorization the shape is
independent of beam polarization. Because of CP invariance and the
Majorana character the energy spectra of $\ell^{-}$ and $\ell^{+}$ are
identical \cite{petcov}.
\vspace{-.5cm}
\section{Summary and conclusions}
\vspace{-.3cm}
We have studied the production
 of neutralinos, $e^{+}+e^{-}\to
\tilde{\chi}^0_1 + \tilde{\chi}^0_2$ with polarized beams and the
subsequent leptonic decay, $\tilde{\chi}^0_2 \to
\tilde{\chi}^0_1 +\ell^{+} + \ell^{-}$. We have fully taken into account the spin
correlations between production and decay. The lepton angular
and energy distributions, and the distribution of the opening angle between the
outgoing leptons at $\sqrt{s} = 500$~GeV have been computed for two representative scenarios. 

The quantum mechanical interference effects modify appreciably the decay lepton angular
distributions. Especially for gaugino-like neutralinos we find contributions up
to
$15\%$ from spin correlations between production and decay. The shape of the angular
distributions depends on the beam polarization. The effect of spin correlations is strong
for right-handed electrons and gaugino--like neutralinos. It is smaller
for left-handed electrons in both scenarios considered.
The mixing character of the neutralinos influences the beam polarization dependence of
the total cross section. Therefore, the
polarization of both beams is very useful for constraining
the parameters of the MSSM.

Owing to the Majorana
character of the neutralinos the quantum mechanical interference effects between various
polarization states of the decaying neutralino
$\tilde{\chi}^0_2$ cancel in the energy spectrum and in the
distribution of the $\ell^{-} \ell^{+}$-opening angle. Consequently,
 these distributions factorize into production and decay similarly to the case of
spinless particles. Accordingly, the shape
 of these distributions is independent of the beam polarization.

\vspace{.5cm}
G.~M.-P.\ thanks M.~Jezabek and the other organizers of the Epiphany
Conference for the friendly atmosphere during the Conference.
We are grateful to V.~Latussek for his support in the development of
the numerical program. This work was also supported by the `Fond zur
F\"orderung der wissenschaftlichen Forschung' of Austria, Project
No. P10843-PHY.



\begin{picture}(9,5)
\put(0,0){\includegraphics{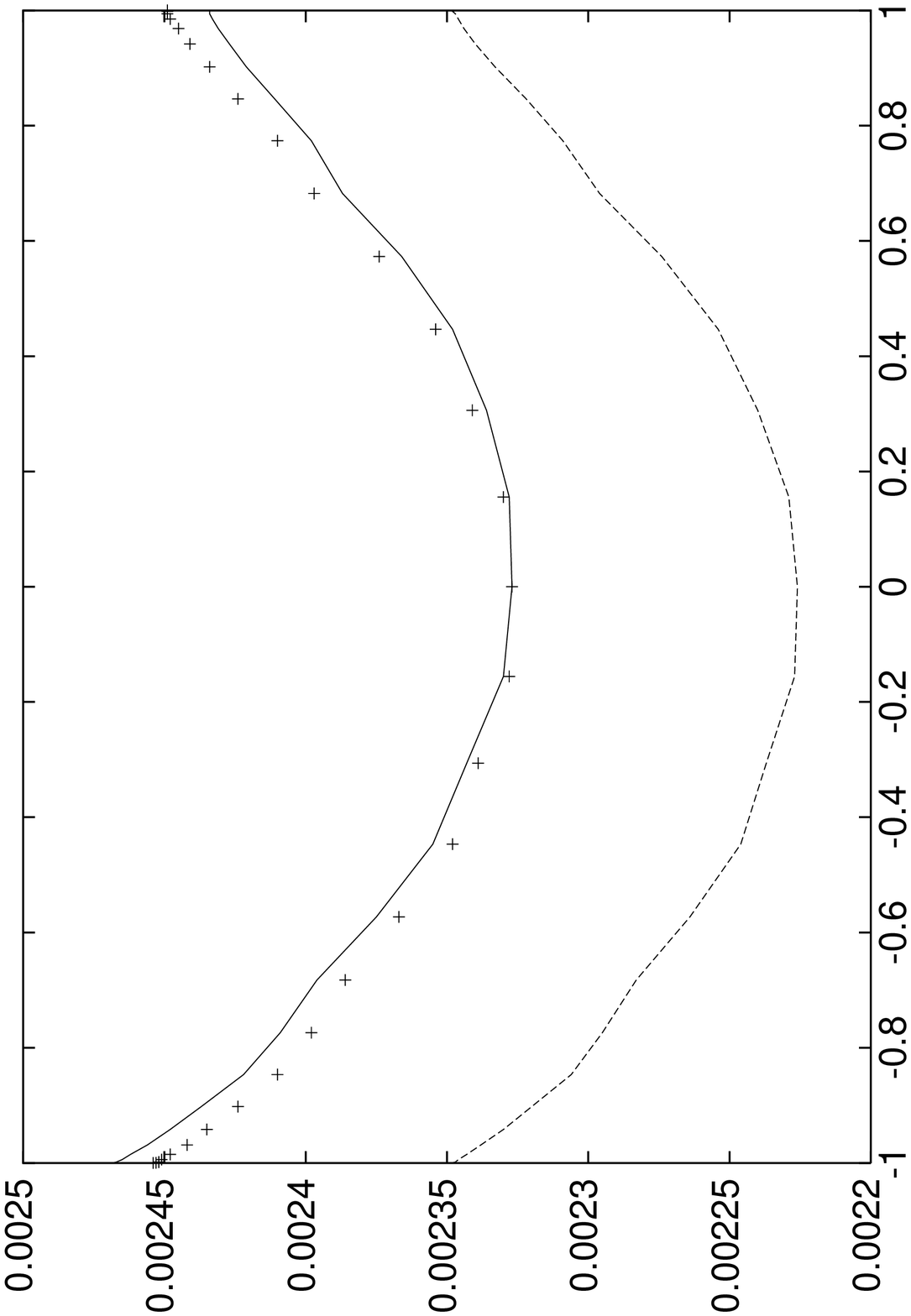}}
\put(130,-120){$ \scriptscriptstyle\cos\Theta_{-}$}
\put(-10,0){$ \scriptscriptstyle\frac{d\sigma}{d\Theta_{-}}/pb$}
\label{w1}
\put(-10,-148)
{\parbox{5.7cm}{\scriptsize Fig.~1: Lepton angular distribution in (A) for
 $m_0=80$ GeV with spin correlations (upper solid),
for assumed factorization (dotted) and for pure $Z^0$-exchange (lower solid).}}
\put(12,0){\includegraphics{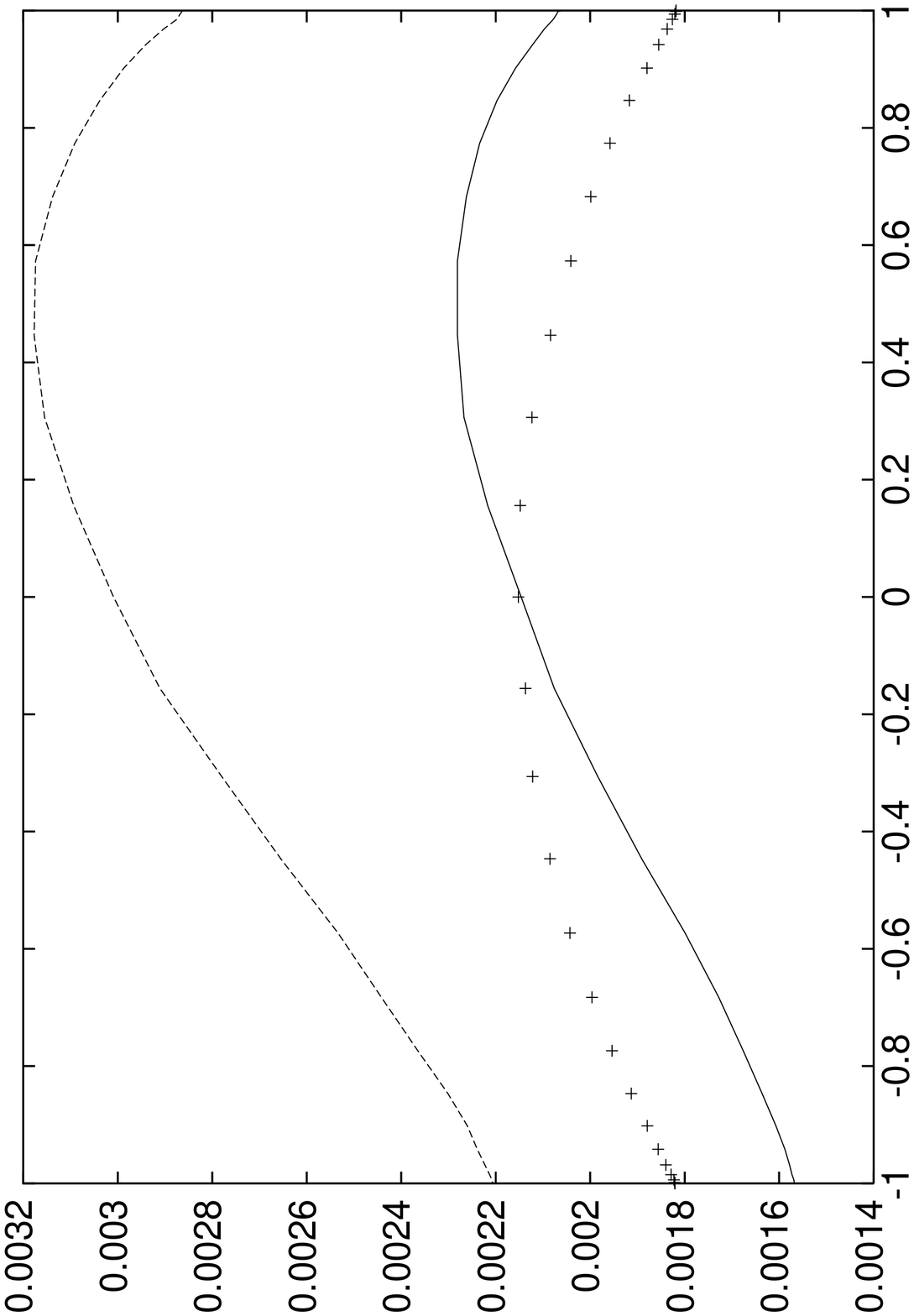}}
\put(310,-120){$\scriptscriptstyle\cos\Theta_{-}$}
\put(170,0){$ \scriptscriptstyle\frac{d\sigma}{d\Theta_{-}}/pb$}
\end{picture}
\label{80w2}
\put(170,-148)
{\parbox{5.7cm}{\scriptsize Fig.~2: Lepton angular distribution in (B)
for $m_0=280$ GeV
with spin correlations (lower solid), for assumed
 factorization (dotted) and for pure slepton-exchange (upper solid).}}


\begin{picture}(10,5)
\put(0,0){\includegraphics{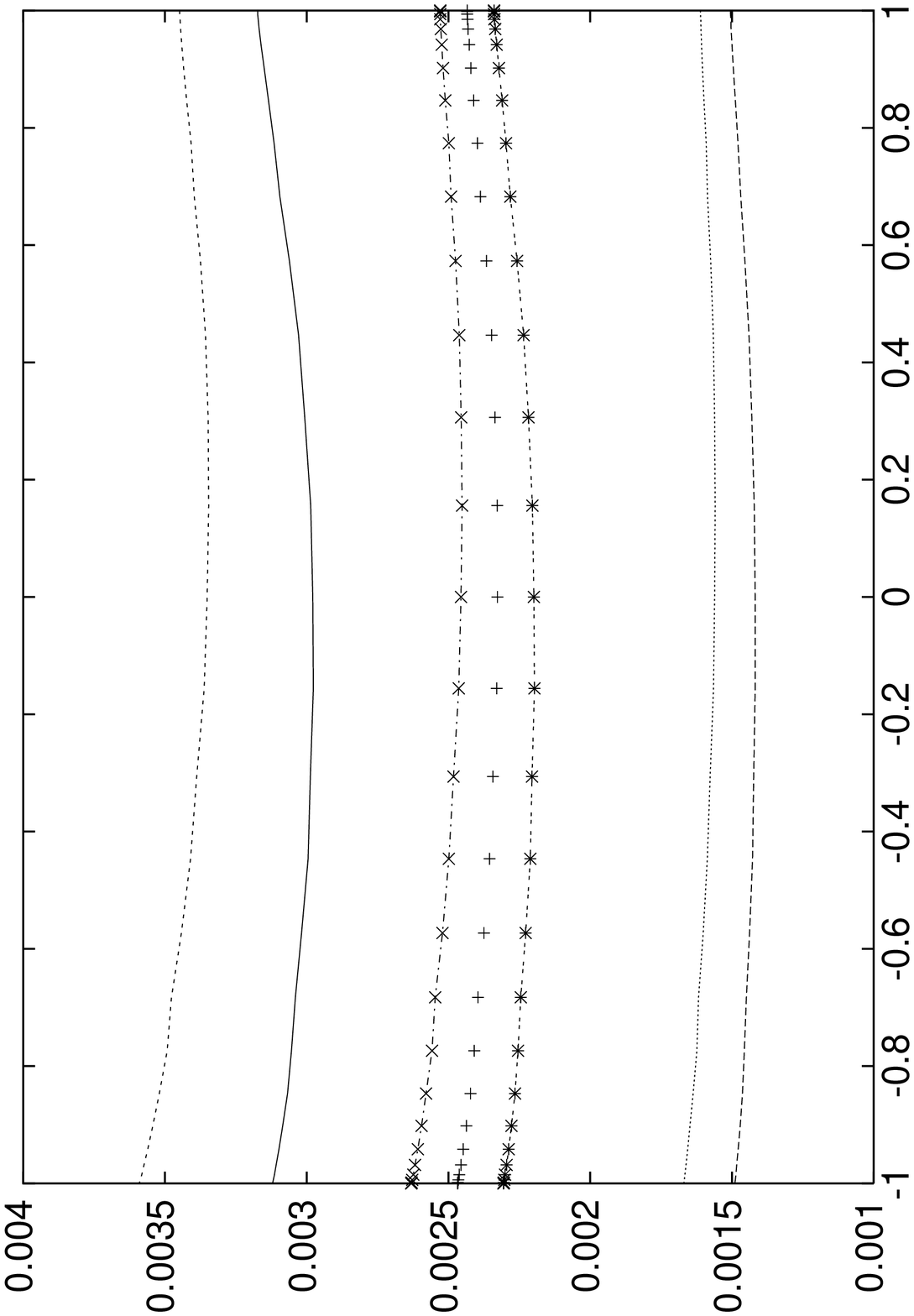}}
\put(130,-120){$\scriptscriptstyle \cos\Theta_{-}$}
\put(-10,0){$ \scriptscriptstyle \frac{d\sigma}{d\Theta_{-}}/pb$}
\label{200w2}
\put(-10,-148)
{\parbox{5.7cm}{\scriptsize Fig.~3: Lepton angular distribution in (A)
 with beam polarization. Order of lines according to the order of total
 cross sections in Table~1.}}
\put(12,0){\includegraphics{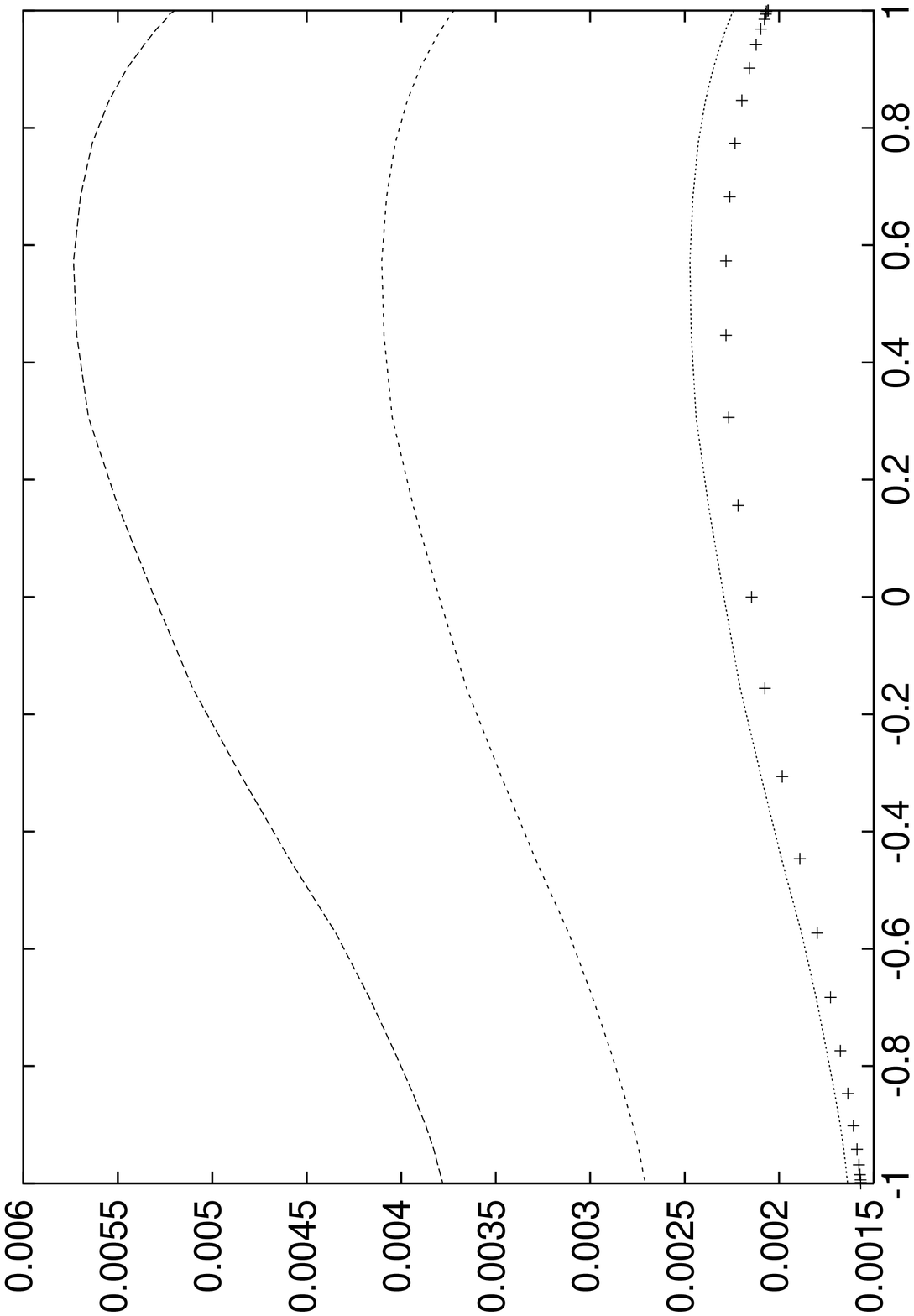}}
\put(310,-120){$ \scriptscriptstyle\cos\Theta_{-}$}
\put(170,0){$ \scriptscriptstyle\frac{d\sigma}{d\Theta_{-}}/pb$}
\label{80w3}
\put(185,-148)
{\parbox{5.7cm}{\scriptsize Fig.~4:
Lepton angular distribution in (B) for left-handed polarized
electrons, ($-+$),($-0$),($--$) (solid) and for unpolarized beams
(dotted). Order of lines according to the order of total cross
sections in Table~1.}}
\end{picture}

\vspace{6cm}
\begin{picture}(10,5)
\put(0,0){\includegraphics{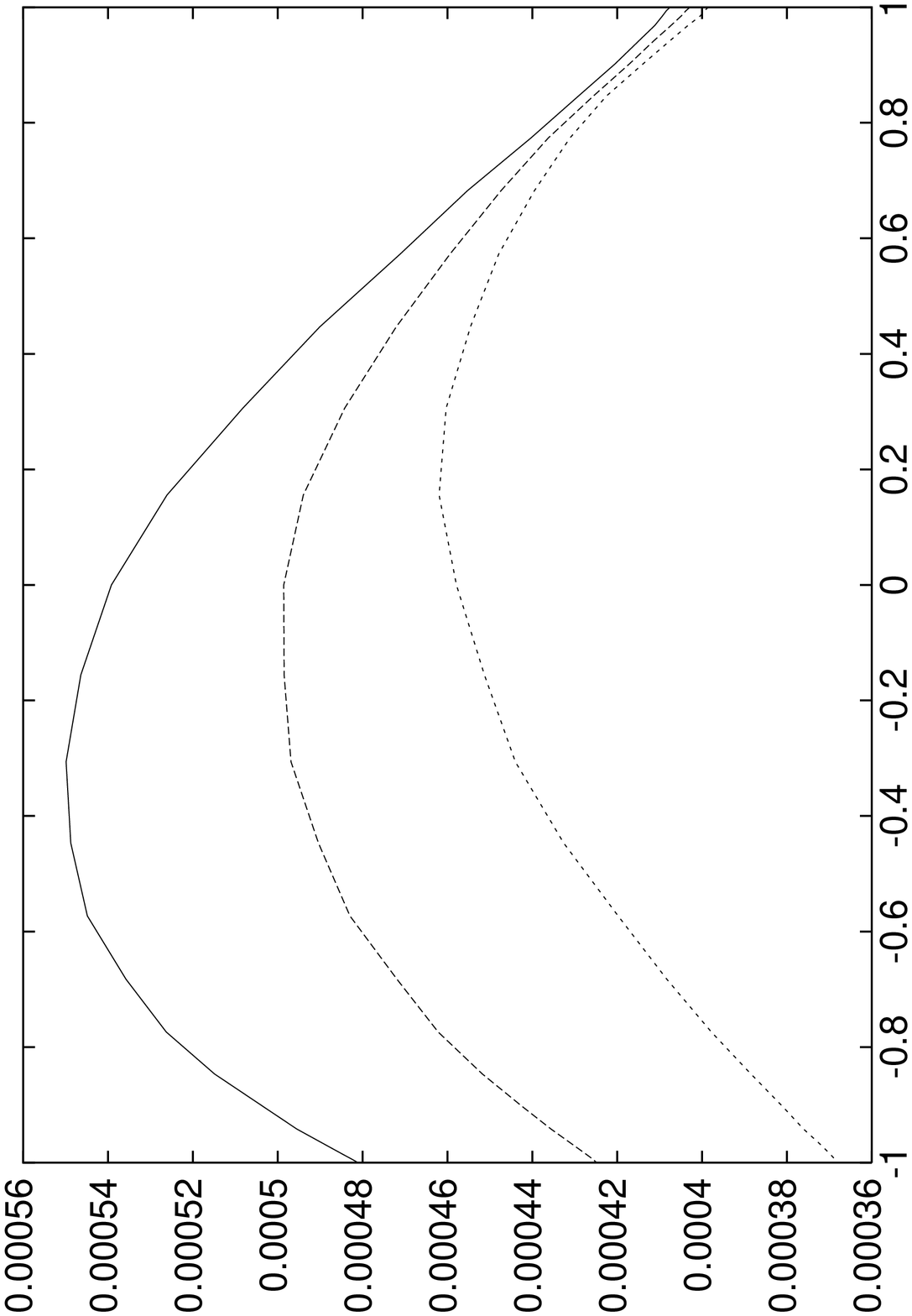}}
\put(130,-120){$ \scriptscriptstyle\cos\Theta_{-}$}
\put(-10,0){$ \scriptscriptstyle\frac{d\sigma}{d\Theta_{-}}/pb$}
\put(-10,-156)
{\parbox{5.7cm}{\scriptsize Fig.~5:
Lepton angular distribution in (B) for right-handed polarized
                       electrons, ($+-$),($+0$),($++$) (solid).
Order of lines according to the order of total cross sections in Table~1.}}
\put(12,0){\includegraphics{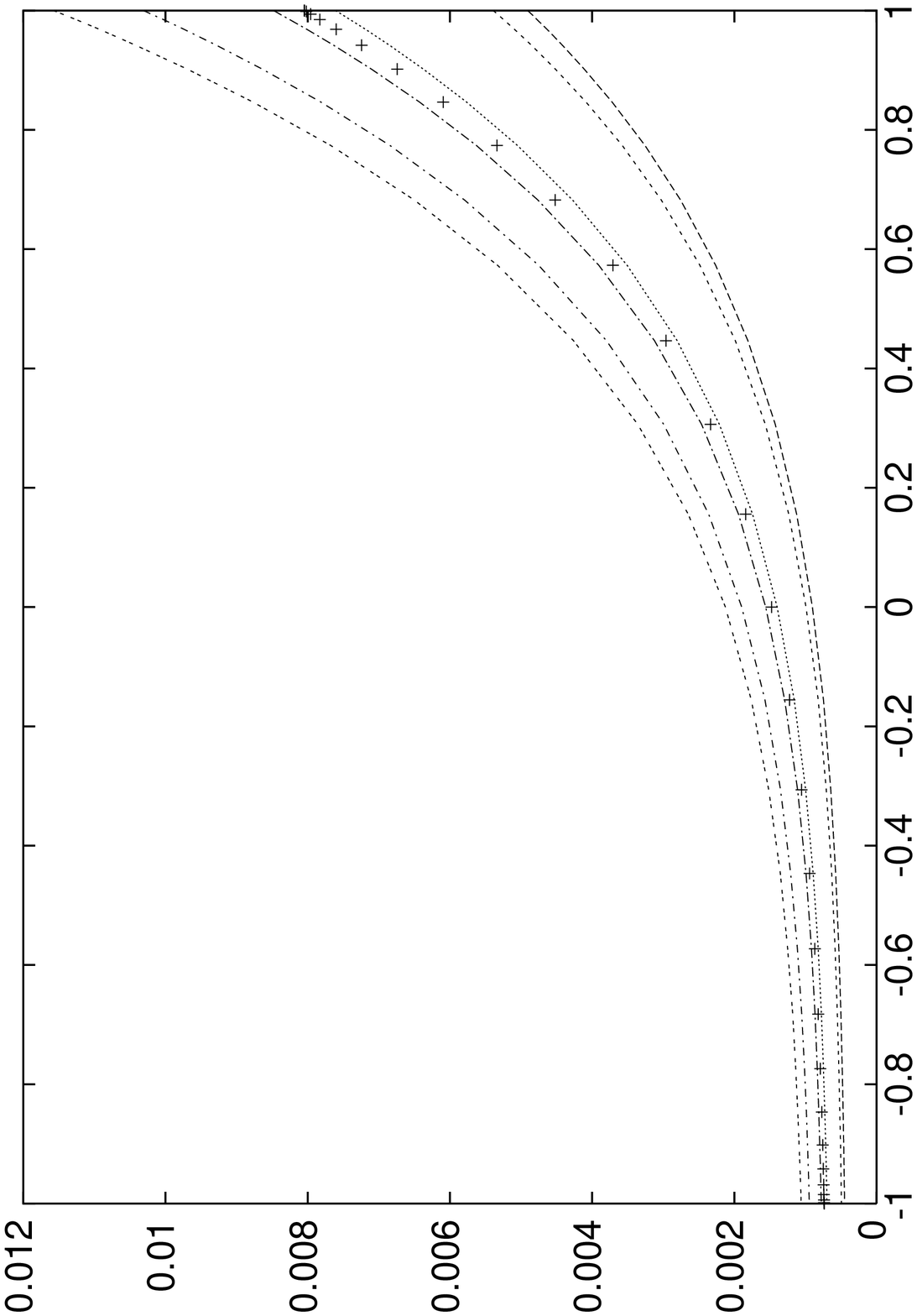}}
\put(310,-120){$ \scriptscriptstyle\cos\Theta_{+-}$}
\put(170,0){$ \scriptscriptstyle\frac{d\sigma}{d\Theta_{+-}}/pb$}
\put(185,-156)
{\parbox{5.7cm}{\scriptsize Fig.~6:
Opening angle distribution  in (A) with beam polarization. Order of
                       lines according to
                       the order of total cross sections in Table~1
                       (solid) and for unpolarized case (dotted).}}
\end{picture}
\label{zw2}

\vspace{6cm}
\begin{picture}(10,5)
\put(0,0){\includegraphics{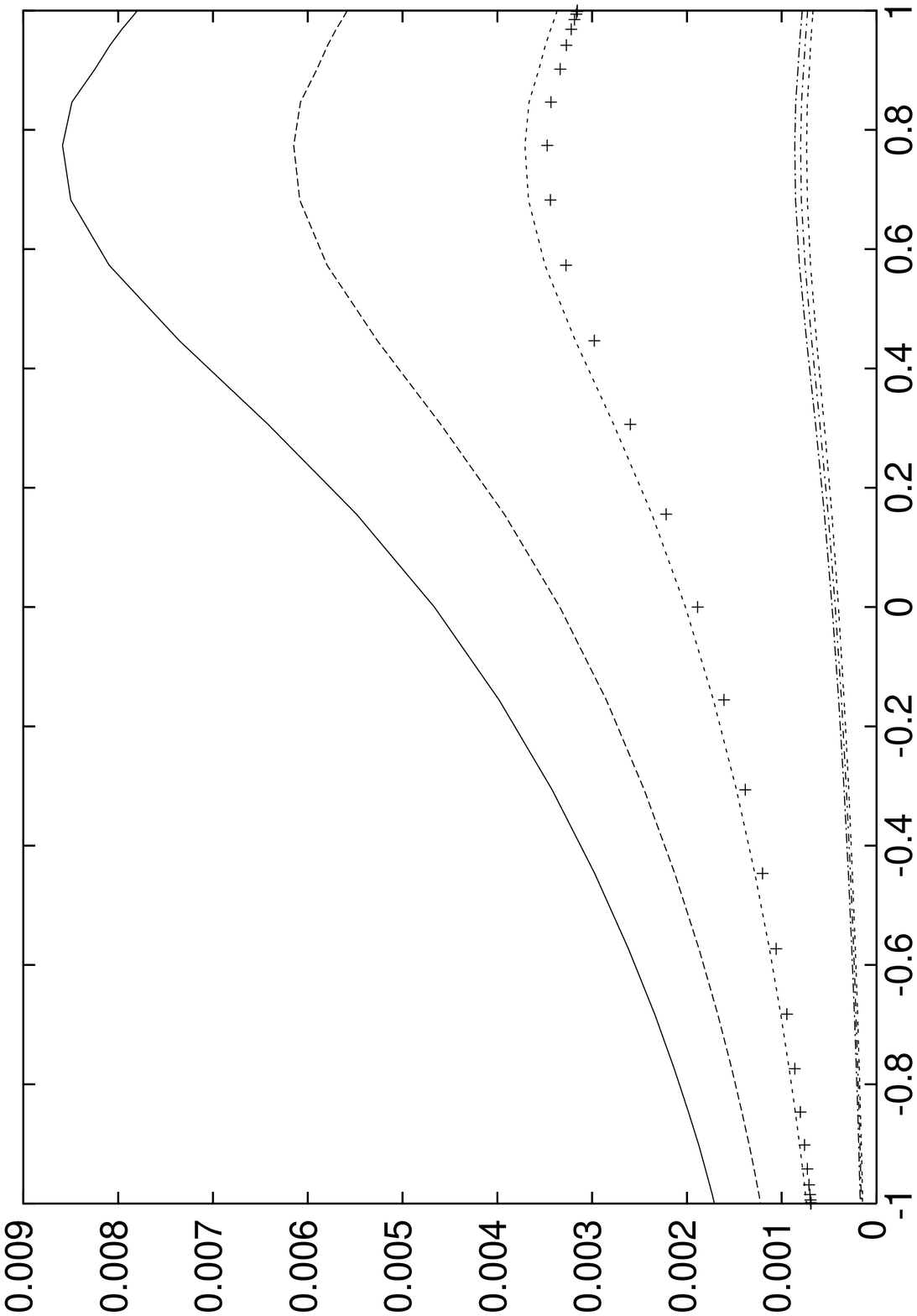}}
\put(130,-120){$ \scriptscriptstyle\cos\Theta_{+-}$}
\put(-10,0){$ \scriptscriptstyle\frac{d\sigma}{d\Theta_{+-}}/pb$}
\label{80zw3}
\put(-10,-148)
{\parbox{5.7cm}{\scriptsize Fig.~7:
Opening angle distribution in (B) with beam polarization. Order of
lines according to the order of total cross sections in Table~1
(solid) and for unpolarized beams (dotted).}}
\put(12,0){\includegraphics{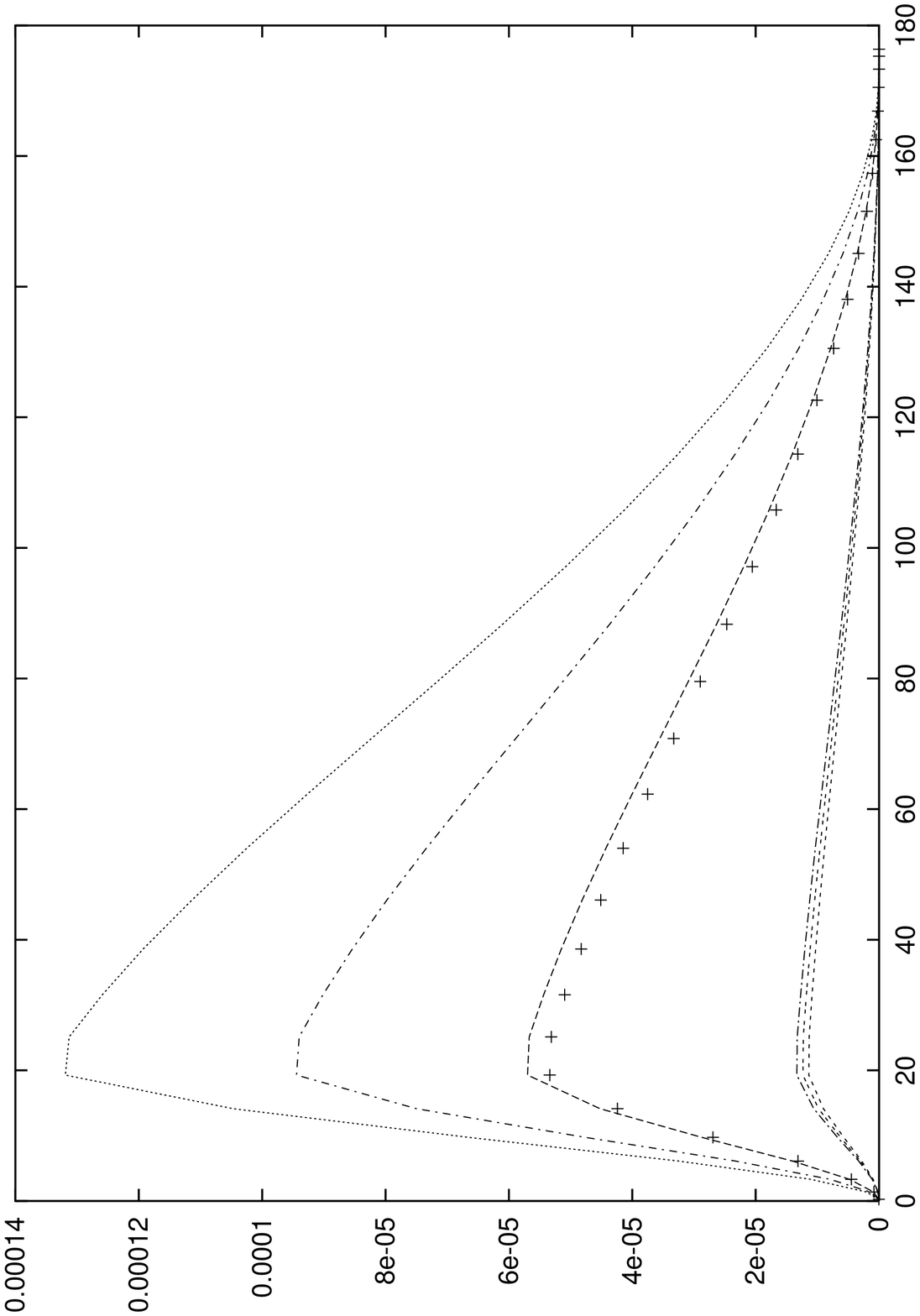}}
\put(310,-120){$ \scriptscriptstyle E_{-}/GeV$}
\put(170,0){$ \scriptscriptstyle \frac{d\sigma}{d E_{-}}$}
\put(185,-143)
{\parbox{5.7cm}{\scriptsize Fig.~8:
Energy distribution in (B) with beam polarization. Order of lines
according to the
order of total cross sections in Table~1 (solid) and for
unpolarized beams (dotted).}}
\end{picture}
\label{e1}

\end{document}